\documentstyle[aps,prl,preprint,floats,epsfig]{revtex}

\begin{document}

\preprint{\tighten\vbox{\hbox{\hfil CLNS 00/1697}
                        \hbox{\hfil CLEO 00-21}
}}

\title{Bounds on the CP Asymmetry in $b \rightarrow s \gamma$ Decays}  

\author{CLEO Collaboration}
\date{January 2, 2001}

\maketitle
\tighten

\begin{abstract} 
We have measured the CP asymmetry ${\cal A}_{CP} \equiv (\Gamma(b
\rightarrow s \gamma) - \Gamma(\bar b \rightarrow \bar s
\gamma))/(\Gamma(b \rightarrow s \gamma) + \Gamma(\bar b \rightarrow
\bar s \gamma))$ to be ${\cal A}_{CP} = (-0.079 \pm 0.108 \pm
0.022)(1.0 \pm 0.030)$, implying that, at 90\% confidence level,
${\cal A}_{CP}$ lies between --0.27 and +0.10.  These limits rule out
some extreme non-Standard-Model predictions, but are consistent with
most, as well as with the Standard Model.  \end{abstract} \newpage

{
\renewcommand{\thefootnote}{\fnsymbol{footnote}}

% Insert author and address list here
\begin{center}
T.~E.~Coan,$^{1}$ V.~Fadeyev,$^{1}$ Y.~Maravin,$^{1}$
I.~Narsky,$^{1}$ R.~Stroynowski,$^{1}$ J.~Ye,$^{1}$
T.~Wlodek,$^{1}$
M.~Artuso,$^{2}$ R.~Ayad,$^{2}$ C.~Boulahouache,$^{2}$
% ...
\end{center}
 
\small
\begin{center}
$^{1}${Southern Methodist University, Dallas, Texas 75275}\\
$^{2}${Syracuse University, Syracuse, New York 13244}\\
% ...
\end{center}

\setcounter{footnote}{0}
}
\newpage

Direct CP violation can lead to a difference between the rates for
$b \rightarrow s \gamma$ and $\bar b \rightarrow \bar s \gamma$, giving rise to
a non-zero value for the CP asymmetry
$$ {\cal A}_{CP} \equiv \frac
{\Gamma (b \rightarrow s \gamma ) - \Gamma (\bar b \rightarrow \bar s \gamma )}
{\Gamma (b \rightarrow s \gamma ) + \Gamma (\bar b \rightarrow \bar s \gamma )}
\ \ .$$
\noindent Such an asymmetry occurs only if the decay is due to two or
more amplitudes with differing strong phases and differing weak
phases. The Standard Model (SM) predicts\cite{theory_SM} that this
asymmetry is very small, less than 1\%.  Recent theoretical
work\cite{theory_SM,theory_nonSM} suggests that non-SM physics may
contribute significantly to a CP asymmetry, giving asymmetries perhaps
as large as 10 -- 40\%.  The advantage of an inclusive measurement, in
contrast to an exclusive measurement, {\it e.g.} $B \rightarrow K^*(890)
\gamma$\cite{Jaffe-Prell}, is that the strong phase is calculable.  In
this Letter, we describe a measurement of a CP asymmetry in $B
\rightarrow X_s
\gamma$, a close approximation to ${\cal A}_{CP}$ as defined above, with a small
admixture from $b \rightarrow d \gamma$.
We rule out very large asymmetry values.

The data used in this analysis were taken with the CLEO detector at the Cornell
Electron Storage Ring (CESR), a symmetric $e^+ e^-$ collider.
They consist of 9.1 ${\rm fb}^{-1}$ on the
$\Upsilon$(4S) resonance, and 4.4 ${\rm fb}^{-1}$ at a center-of-mass energy
$\sim$ 60 MeV below the resonance.  The on-resonance sample contains 10 million
$B \bar B$ events and 30 million continuum events, while the off-resonance
sample contains 15 million continuum events.

The CLEO detector\cite{CLEO detector}  measures charged particles over 95\% of
$4\pi$ steradians with a system of cylindrical drift chambers.  (For 2/3 of the
data used here, the innermost tracking chamber was a 3-layer silicon vertex
detector.)   Its barrel and endcap CsI electromagnetic calorimeters cover 98\%
of 4$\pi$.  The energy resolution for photons near 2.5 GeV in the central
angular region, $\vert \cos \theta_\gamma \vert < 0.7$, is 2\%.  Charged
particles are identified by specific ionization measurements ($dE/dX$) in the
outermost drift chamber, and by time-of-flight counters (ToF) placed just
beyond the tracking volume.  Muons are identified by their ability to
penetrate the iron return yoke of the magnet.

The signature for $b \rightarrow s \gamma$ is a photon with energy
sufficiently high that it is unlikely to come from other $B$ decay
processes. We take our photon energy range as 2.2 -- 2.7 GeV. With
this requirement, there is little background from other $B$ decay
processes.  There is substantial background from continuum processes,
and so continuum suppression is a key ingredient in this analysis, as
it is in any $b \rightarrow s \gamma$ analysis.  The other ingredient
needed for an asymmetry measurement is flavor tagging.  Not needed for
an asymmetry measurement, unlike absolute rate measurements of $b
\rightarrow s \gamma$, is a good knowledge of the efficiency, since it
cancels when the ratio is taken.

We use two different methods of flavor tagging.  In the first, we accept events
with a high energy photon and a high momentum  lepton.  The lepton charge tags the
flavor of the `other' $B$, tagging the flavor of $b \rightarrow s \gamma$
correctly 89\% of the time.
$B^0 - \bar B^0$ mixing is the dominant reason for mistags.  The
second method of flavor tagging is `pseudoreconstruction', described below and
used in our previous $b \rightarrow s \gamma$ analysis\cite{bsg_PRL}.  Here, we
are directly tagging the flavor of the $b \rightarrow s \gamma$ decay, so mixing
does not cause mistags.  With
aggressive use of particle identification, we have achieved a correct flavor
identification rate of better than 90\%.

We use Monte Carlo simulation (along with
the measured value for $B^0 - \bar B^0$ mixing) to determine the performance of
our flavor identification methods.  For this, we describe the
$b \rightarrow s \gamma$ process with a spectator model\cite{Ali}, and vary
the spectator
model parameters to determine systematic errors to flavor identification from
the modelling of the $b \rightarrow s \gamma$ process.

Pseudoreconstruction was originally introduced as a continuum background
suppression technique.  Requiring a high momentum lepton also suppresses the
continuum, particularly when use is made of the angle between lepton and
photon.  So, our flavor-tagging methods are also part of our continuum
suppression.

We select hadronic events having normalized Fox-Wolfram\cite{Fox-Wolfram}  
second moment $R_2$
less than 0.45, and containing a photon with energy between 2.2 and 2.7 GeV, in
the ``good barrel'' region of the calorimeter ($\vert \cos \theta_\gamma \vert <
0.7$).  The high energy photon must not form a $\pi^0$ or $\eta$ with any other
photon in the event.

Part of our continuum suppression comes from eight carefully chosen event shape
variables: $R_2$, $S_\perp$ (a measure of the momentum transverse to the photon
direction\cite{Jesse_thesis}), $R^\prime_2$ (the value of $R_2$ in the primed frame,
the rest frame of $e^+ e^-$ following an assumed initial state radiation of the
high energy photon, with $R_2$ evaluated excluding the photon),
$\cos \theta^\prime$ ($\theta^\prime$ the angle, in the primed frame, between
the photon and the thrust axis of the rest of the event), and the energies in
$20^\circ$ and $30^\circ$ cones,
parallel and antiparallel to the high energy photon direction.  While no
individual variable has strong discrimination power, each posesses some.
Consequently, we combine the eight variables into a single variable $r$ that
tends towards +1 for $b \rightarrow s \gamma$ events and tends towards --1 for
continuum background events, using a neural
network\cite{Jesse_thesis}. Distributions in the neural net variable
$r$, for Monte Carlo samples of $b \rightarrow s \gamma$ signal and
continuum background, and off resonance data, are shown in ref.~\cite{bsg_PRL}.

While the background to $b \rightarrow s \gamma$ from other $B$ decay
processes is small, it is not negligible.  We investigated it with a
$B \bar B$ Monte Carlo sample that included contributions from $b
\rightarrow u$ and $b \rightarrow s g$ processes, as well as the
dominant $b \rightarrow c$ decay.  We found that the overwhelming
source of background ($\sim$90\%) is photons from $\pi^0$ or $\eta$
decay.  Consequently we tuned the Monte Carlo to match the data in
$\pi^0$ and $\eta$ yields. 

For the lepton tagging method of flavor identification, we require a high
momentum lepton ($1.4 < P < 2.2$ GeV/$c$), either a muon (identified by passing
through at least 5 interaction lengths of material) or an electron (identified
by $E/P$, track-cluster matching, $dE/dX$, and shower shape).  The distribution
in cosine of angle between lepton and high energy photon, $\cos \theta_{\ell -
\gamma}$, is isotropic for $b \rightarrow s \gamma$ signal events, and strongly
back-to-back peaked for continuum background events.  We make loose cuts on $r$
and $\cos \theta_{\ell - \gamma}$, and then define weights $w_i$ in the 2D
$r - \cos \theta_{\ell - \gamma}$ space, where
$w_i = s_i/[s_i + (1 + a)b_i]\ ,$ 
$s_i$ is the expected signal yield in the i-th bin, $b_i$ is the
expected continuum background yield in that bin, and $a$ is the luminosity scale
factor between on-resonance and off-resonance data samples ($\approx 2.0$).
Weights so defined minimize the statistical error on the $b \rightarrow s
\gamma$ yields. Expected yields are obtained from Monte Carlo
simulation. Should the Monte Carlo simulation of signal or background
be flawed, the weights will not be optimum, but they will not lead to
incorrect results.
We sum weights on and off resonance, and subtract the off-resonance sum,
scaled by ${\cal L}/E^2_{cm}$, from the on-resonance sum (the On-Off
subtraction).  We also subtract the
background from $B$ decay processes.  We do this separately
for events tagged with an $\ell^-$ and events tagged with an $\ell^+$, and
compute the raw asymmetry, obtaining ${\cal A}^{raw}_{lep} = 0.148 \pm 0.141$.
Summed weights, and the number of events with non-zero weight, are given in
Table~\ref{tab.Data-yields-Lepton}.

\begin{table}[h]
\begin{center}
\begin{tabular}{|c|ccc|}
\hline
  &           $N$       &     $W(b)$        &    $W(\bar b)$      \\
\hline
On          & 507       & 127.10 $\pm$  8.90 & 107.34 $\pm$  7.81  \\ 
Off         & 135       &  24.96 $\pm$  3.25 &  23.24 $\pm$  3.37  \\ 
Sca         & 279.9     &  51.73 $\pm$  6.76 &  48.08 $\pm$  6.97  \\ 
$B \bar B$  &  39.8     &       12.80        &        12.80        \\ 
\hline
Sub  & 187.3 &  62.57 $\pm$ 11.18 &  46.46 $\pm$ 10.47 \\ 
\hline
\end{tabular}
\end{center}
\caption{Yields (weights) for $b$-flavored ($W(b)$), and
$\bar b$-flavored ($W(\bar b)$), for On, Off,
Scaled Off, $B \bar B$ background, and On -- Scaled Off -- $B \bar B$
background.  The number of events passing all cuts, $N$, is also given.
Lepton tag analysis.}
\label{tab.Data-yields-Lepton}
\end{table}

This raw asymmetry must be corrected for the mistag fraction $\alpha$, by
dividing it by $1 - 2\alpha$.  
Mistags come mainly from $B^0 - \bar B^0$ mixing, a contribution  to $\alpha$
of $\chi_d$/2, which we take from CLEO's dilepton mixing
measurement\cite{mixing-CLEO}, $\chi_d = 0.157$.  (By using CLEO's dilepton
mixing measurement, we eliminate some systematic errors, in particular the
uncertainty in the ratio of $B^+ B^-$ to $B^0 \bar B^0$ events.) Other sources
include secondary decays $b \rightarrow c \rightarrow s \ell \nu$, leptons from
$J/\psi$ decays, and hadrons misidentified as leptons.  These we determine, from
Monte Carlo studies, to be 0.033.  Adding this to the mixing contribution gives
$\alpha = 0.112 \pm 0.013$.  The corrected value of asymmetry is
$${\cal A}_{lepton\ tag} = +0.191 \pm 0.181\ \ .$$

For events not containing high momentum leptons, we use the
pseudoreconstruction method of flavor identification.  We search those
events for combinations of particles that
reconstruct to a $B \rightarrow X_s \gamma$ decay.  For $X_s$ we use a
$K^0_S \rightarrow \pi^+ \pi^-$ or a charged track consistent with a $K^\pm$;
and 1 -- 4 pions, of which at most one may be
a $\pi^0$.  We calculate the candidate $B$ momentum $P$, energy $E$, and
beam-constrained mass $M \equiv \sqrt{E^2_{beam} - P^2}$.  A reconstruction
is deemed acceptable if it has $\chi^2_B < 20$, where
$$\chi^2_B \equiv \left( \frac{E - E_{beam}}{\sigma_E}\right)^2 + \left( \frac
{M - M_B}{\sigma_M}\right)^2\ \ .$$
\noindent (Typically, $\sigma_E \approx 46$ MeV, $\sigma_M \approx 3.5$ MeV.)  
If an event contains more than one acceptable reconstruction, the
``best'' is selected on the basis of an overall $\chi^2$ consisting of
$\chi^2_B$ plus contributions from particle identification.

We discriminate between signal and background using $\chi^2_B$, $r$ (from the
shape variables), and $\vert \cos \theta_{tt} \vert$, where $\theta_{tt}$ is the
angle between the thrust axis of the candidate $B$ and the thrust axis of the
rest of the event.  These three variables are combined into a single variable
$r_c$, that tends towards $+1$ for signal and $-1$ for continuum background, using
a neural net.  As for the lepton tag analysis, we count weights, not events,
with weights defined as earlier, now a function of $r_c$.  Weights are summed
for events tagged as $b$-flavored, $\bar b$-flavored, and ambiguous, separately
on and off resonance, and an On-Off resonance subtraction is performed.

Note that with pseudoreconstruction the flavor can be correctly or incorrectly
tagged (if it reconstructs to a charged $B$, or to a neutral $B$ decaying to
$K^\pm$), or it can be tagged as ambiguous (if it reconstructs to a neutral $B$
decaying to $K^0_S$).  A major source of incorrect tags is using a $\pi^\pm$ for
a $K^\pm$.  To reduce this source of incorrect tags, we have been aggressive in
our $K^\pm$ identification.

For runs for which the ToF measurement was available and reliable, 3/4
of the luminosity, we used both ToF and $dE/dX$ for $K^\pm$ and
$\pi^\pm$ identification; for the other 1/4 of the luminosity, we used
only $dE/dX$.  The cut when both ToF and $dE/dX$ were used was a
$\Delta \chi^2$ cut, the difference between the $\chi^2$ of the ToF
plus $dE/dX$ fits assuming $K^\pm$ and assuming $\pi^\pm$.  Whether or
not ToF was used, we used a 3$\sigma$ cut (or the 2D equivalent) for
$\pi^\pm$ identification.  For $K^\pm$, if the $K - \pi$ separation
was very good ($>6\sigma$), or if it was hopeless ($\sim 0\sigma$), we
used a 3$\sigma$ cut or the equivalent.  In between, we cut more
harshly, in a manner that depended on the computed $K - \pi$
separation.  We achieved a misidentification probability of 8.5\% when
ToF and $dE/dX$ were used, 12.2\% when only $dE/dX$ was used.

Because the pseudoreconstruction analysis has three possible outcomes -- 
$b$-flavored, $\bar b$-flavored, ambiguous  -- the formulation is more complex
than the lepton tag case.  Let $\alpha$ be the probability that a taggable event
be incorrectly tagged, $\beta$ be the probability that a taggable event be
declared ambiguous, and $\gamma$ be the probability that an event which is
actually ambiguous be tagged as $b$ or $\bar b$.  For  the analysis using ToF
and $dE/dX$, we use Monte Carlo to obtain $\alpha = 0.085 \pm 0.006,
\ \beta = 0.016 \pm 0.004,\ \gamma = 0.40 \pm 0.05$, while for the analysis
with $dE/dX$ only, we obtain
$\alpha = 0.122 \pm 0.007,\  \beta = 0.013 \pm 0.005,\ \gamma = 0.49 \pm 0.06$.

Using $\alpha, \beta, \gamma$, we compute a raw asymmetry
$$ {\cal A}^{raw}_{pseudo} = \frac{N(b) - N(\bar b)}{N(b) + N(\bar b) -
(\gamma/(1-\gamma)) N(ambig)}\ ,$$
\noindent and then correct by a factor
$(1 - \beta/(1 - \gamma))/(1 - 2\alpha - \beta)$.  This factor is more
complicated than $1 -2\alpha$, but is essentially equivalent to it.  We combine
the corrected asymmetries for runs with ToF and $dE/dX$ and runs with $dE/dX$
only, weighting each by the expected statistical accuracy.  We find
$$ {\cal A}_{pseudo} = -0.178 \pm 0.132 \ \ .$$
Summed weights, and the number of events with non-zero weight, are
given in Table~\ref{tab.Data-yields}.

\begin{table}[h]
\begin{center}
\begin{tabular}{|c|cccc|}
\hline
        &  $N$   &      $W(b)$       &   $W(\bar b)$     &      $W(?)$      \\
\hline
On      & 5542   & 171.17 $\pm$ 6.81 & 174.73 $\pm$ 6.97 & 22.97 $\pm$ 2.72 \\ 
Off     & 2318   &  52.80 $\pm$ 3.11 &  48.37 $\pm$ 2.88 &  5.49 $\pm$ 0.96 \\ 
Sca     & 4877.8 & 111.55 $\pm$ 6.56 & 101.50 $\pm$ 6.06 & 11.50 $\pm$ 2.02 \\ 
$B \bar B$ & 113.2 &       8.67      &         8.67      &        1.20      \\ 
\hline
Sub     &  551.0 &  56.95 $\pm$ 9.46 &  64.56 $\pm$ 9.23 & 10.27 $\pm$ 3.39 \\ 
\hline
\hline
On      & 2408   &  65.49 $\pm$ 3.81 &  72.34 $\pm$ 4.26 &  8.21 $\pm$ 1.34 \\ 
Off     & 1062   &  24.18 $\pm$ 2.13 &  20.63 $\pm$ 1.85 &  2.69 $\pm$ 0.70 \\ 
Sca     & 2113.7 &  47.52 $\pm$ 4.19 &  40.70 $\pm$ 3.68 &  5.53 $\pm$ 1.44 \\ 
$B \bar B$ &  34.6 &       2.93      &         2.93      &        0.41      \\ 
\hline
Sub     &  259.7 &  15.03 $\pm$ 5.67 &  28.71 $\pm$ 5.63 &  2.27 $\pm$ 1.97 \\ 
\hline
\end{tabular}
\end{center}
\caption{Yields (weights) for $b$-flavored ($W(b)$),
$\bar b$-flavored ($W(\bar b)$), and ambiguous ($W(?)$), for On, Off,
Scaled Off, $B \bar B$ background, and On -- Scaled Off -- $B \bar B$
background.  The number of events passing all cuts, $N$, is also
given.  The upper half of the table is the pseudoreconstruction
analysis with ToF and $dE/dX$; the lower half is with $dE/dx$
only.}
\label{tab.Data-yields}
\end{table}

Pseudoreconstruction and lepton tag results are consistent with each other, and
are statistically independent.
We combine them, weighting each by the expected statistical accuracy, giving
$$ {\cal A}_{comb} = -0.072 \pm 0.107 \ \ .$$

The increased statistical power from using weights, as compared to
counting all events with non-zero weights, was 1.6 for the lepton tag
analysis and 4.0 for the pseudoreconstruction analysis.

False asymmetries in the lepton tag analysis would be caused by a difference in
detection plus identification efficiency for electrons {\em vs.} positrons, or $\mu^-$
{\em vs.} $\mu^+$.  By measuring the rates for $\ell^-$ and $\ell^+$ from On-Off
subtracted data, we find such false asymmetries to be consistent with zero, and
safely bounded by $\pm0.01$, which we take as the additive systematic error for
the lepton tag analysis.

False asymmetries in the pseudoreconstruction analysis would be caused by
particle identification biases favoring $K^\pm$ over $K^\mp$, or favoring
$\pi^\pm$ over $\pi^\mp$.  By measuring the momentum spectra (On-Off subtracted)
for candidate $K^-$, $K^+$, $\pi^-$, and $\pi^+$, noting the biases in particle
identification that these spectra could accommodate, and translating that into a
limit on false asymmetry, we established that an additive systematic error of
$\pm0.01$ adequately covers the uncertainty from particle identification.

False asymmetries in the pseudoreconstruction analysis would also be caused by
particle detection biases, in particular from the different interaction cross
sections for $K^+$ and $K^-$.  There is about 1 ${\rm g/cm}^2$ of material
between
the interaction point and the back of the tracking volume.  This, with the known
$K^+$ and $K^-$ cross sections, limits that source of false asymmetry to
$\pm0.007$.

If either $B \rightarrow \pi^0 X$ or $B \rightarrow \eta X$ had a nonzero CP
asymmetry, then that would creep into the measured $b \rightarrow s \gamma$
asymmetry through the feeddowns $\pi^0 \rightarrow \gamma$ and
$\eta \rightarrow \gamma$.  In making the subtraction for $B \bar B$ background,
we have so far assumed these asymmetries are zero (as expected theoretically),
apportioning the background equally between $b$ and $\bar b$.  In addition, we
have measured the $\pi^0$ and $\eta$ asymmetries, treating each as if it were
the high energy photon, and following the same method as for measuring the
$b \rightarrow s \gamma$ asymmetry.  Combining the lepton tag and
pseudoreconstruction analyses, we find a $\pi^0$ asymmetry of +0.070 $\pm$
0.056, an $\eta$ asymmetry of +0.156 $\pm$ 0.333, both consistent with zero.
Allowing for the fact that the lepton tag and pseudoreconstruction asymmetries
may be different, we find corrections to the lepton tag analysis of +0.034 $\pm$
0.025, to the pseudoreconstruction analysis of --0.024 $\pm$ 0.017, and to the
combined analysis of --0.007 $\pm$ 0.014.

Multiplicative systematic errors result from uncertainties in the mistag rate,
uncertainties in the scale factor for the off-resonance subtraction
(resulting from change in event shape with the 60 MeV $E_{cm}$ shift), and
uncertainties in the subtraction for other $B$ decay processes. For the
lepton tag analysis, we estimate $\pm 3.4$\% from the mistag uncertainty,
$\pm1.1$\% from the Off-resonance subtraction uncertainty, $\pm 4.0$\% from the
$B \bar B$ subtraction uncertainty, $\pm 5.4$\% total.  For the
pseudoreconstruction analysis, we estimate $\pm1.6$\% from the mistag
uncertainty, $\pm1.9$\% from the Off-resonance subtraction uncertainty,
$\pm2.6$\% from the $B \bar B$ subtraction uncertainty, $\pm3.6$\% total.  The
combined multiplicative systematic error, lepton tag plus pseudoreconstruction,
is  $\pm3.0$\%.

We have verified that the analysis proceedure, when applied to Monte
Carlo samples with actual $CP$ asymmetry, correctly finds the $CP$
asymmetry introduced, whether large, small, or zero.

The asymmetry we have measured is a weighted sum over a variety of $b
\rightarrow s \gamma$ decays -- charged $B$, neutral $B$; low mass
$X_s$, high mass $X_s$; {\em etc.}  In particular, those $b
\rightarrow s \gamma$ decays that are inherently ambiguous under a
pseudoreconstruction analysis, {\em i.e.}, neutral $B$ decays to
neutral kaons, have asymmetries measured only by the lepton tag
analysis.  If the different varieties of $b \rightarrow s \gamma$
decays have asymmetries that differ among themselves by no more than
$\pm0.10$, then the unevenness in our weightings will lead to an
asymmetry that differs from the asymmetry with uniform weightings by
no more than $\pm0.02$.  We have looked for dependence of ${\cal
A}_{CP}$ on $M_{X_s}$ or $E_\gamma$, and within our limited statistics
found none.

Also included in the asymmetry we have measured is a component from
$b \rightarrow d \gamma$.  Within the framework of the Standard Model,
the rate for $b \rightarrow d \gamma$ decays is down by a factor of
$\vert V_{td}/V_{ts} \vert^2 \approx 1/20$, but the asymmetry is up by the
reciprocal of the same factor, {\em i.e.} 20, and of opposite sign.
The ratios of efficiencies for $b \rightarrow d \gamma : b
\rightarrow s \gamma$ are 1.1 for the lepton tag analysis and 0.56 for 
pseudoreconstruction; 0.65 combined. While the
misidentification parameter $\alpha$ for the lepton tag analysis is essentially
the same for $b \rightarrow d \gamma$ as it is for $b \rightarrow s \gamma$, for
the pseudoreconstruction analysis, $\alpha$ for $b \rightarrow d \gamma$ is very
poor, $\approx 0.4$, only slightly better than a random guess ($\alpha = 0.5$).
Thus, with the weightings that we have given to the lepton tag and
pseudoreconstruction analyses, and assuming a ratio of branching fractions for
$b \rightarrow d \gamma$ to $b \rightarrow s \gamma$ of 1/20, 
we have measured a weighted sum of CP asymmetries ${\cal A} = 0.965{\cal A}(b
\rightarrow s \gamma) + 0.02{\cal A}(b \rightarrow d \gamma)\ .$

In conclusion, we have measured a CP asymmetry in $b \rightarrow s \gamma$
plus $b \rightarrow d \gamma$ decays.
Our final result is
$${\cal A}_{CP} = (-0.079 \pm 0.108 \pm 0.022)(1.0 \pm 0.030)\ \ .$$
\noindent The first (and by far the largest) error is statistical; the
second is additive systematic and includes an allowance of $\pm0.020$
for the non-uniform weightings over the various $b \rightarrow s \gamma$ decay
modes; the third is multiplicative systematic.

This measurement implies that, at 90\% confidence level, ${\cal A}_{CP}$ lies
between the limits
$$ -0.27 < {\cal A} < +0.10 \ \ .$$
\noindent These limits rule out some extreme non-Standard-Model predictions,
but are consistent with most, as well as with the Standard Model.  Note that the
analysis reported here uses the same data sample as CLEO's
measurement\cite{Jaffe-Prell} of the CP asymmetry in the exclusive decay
$B \rightarrow K^*(892) \gamma$, and so is not statistically independent of it.

We gratefully acknowledge the effort of the CESR staff in providing us with
excellent luminosity and running conditions.
This work was supported by 
the National Science Foundation,
the U.S. Department of Energy,
the Research Corporation,
the Natural Sciences and Engineering Research Council of Canada, 
the Swiss National Science Foundation, 
the Texas Advanced Research Program,
and the Alexander von Humboldt Stiftung.

\end{document}